\documentclass[conference]{IEEEtran}
\usepackage{mathtools}
\usepackage{multicol}
\usepackage{amsmath,amssymb,cite,multirow,subfigure,amstext}
\usepackage{graphicx}
\usepackage{subfigure}
\usepackage{url}
\usepackage{multirow}

\usepackage[linesnumbered,ruled, vlined]{algorithm2e}
\usepackage{algorithmic}
\usepackage{makecell}
\usepackage{cite}
\usepackage{bm}

\usepackage{xcolor}
\usepackage{booktabs}
\usepackage{threeparttable}
\usepackage{stfloats}
\usepackage[marginal]{footmisc}

\usepackage{pifont}
\newcommand\mynuma[1]{\ifcase#1 \or \ding{172}\or \ding{173}\or
	\ding{174}\or \ding{175}\or \ding{176}\or \ding{177}%
	\or \ding{178}\or \ding{179}\or \ding{180}\or \ding{181}\else *\fi\relax}
\newcommand\mynumb[1]{\ifcase#1 \or \ding{182}\or \ding{183}\or
	\ding{184}\or \ding{185}\or \ding{186}\or \ding{187}%
	\or \ding{188}\or \ding{189}\or \ding{190}\or \ding{191}\else *\fi\relax}

\begin{document}
\bstctlcite{IEEEexample:BSTcontrol}

%
\title{A Computationally Efficient Neural Video Compression Accelerator Based on a Sparse CNN-Transformer Hybrid Network}


\author{\IEEEauthorblockN{Siyu Zhang$^{1}$, Wendong Mao$^{2}$, Huihong Shi$^{1}$ and Zhongfeng Wang$^{1,2}$, \emph{Fellow, IEEE}}\\
	\IEEEauthorblockA{$^{1}$ School of Electronic Science and Engineering, Nanjing University, P. R. China}
	\IEEEauthorblockA{$^{2}$ School of Integrated Circuits, Sun Yat-sen University, P. R. China}
	\IEEEauthorblockA{Email: syzhang@smail.nju.edu.cn, maowd@mail.sysu.edu.cn, shihh@smail.nju.edu.cn, zfwang@nju.edu.cn}
}

\maketitle

\begin{abstract}

Video compression is widely used in digital television, surveillance systems, and virtual reality. Real-time video decoding is crucial in practical scenarios. Recently, neural video compression (NVC) combines traditional coding with deep learning, achieving impressive compression efficiency. Nevertheless, the NVC models involve high computational costs and complex memory access patterns, challenging real-time hardware implementations. To relieve this burden, we propose an algorithm and hardware co-design framework named NVCA for video decoding on resource-limited devices. Firstly, a CNN-Transformer hybrid network is developed to improve compression performance by capturing multi-scale non-local features. In addition, we propose a fast algorithm-based sparse strategy that leverages the dual advantages of pruning and fast algorithms, sufficiently reducing computational complexity while maintaining video compression efficiency. Secondly, a reconfigurable sparse computing core is designed to flexibly support sparse convolutions and deconvolutions based on the fast algorithm-based sparse strategy. Furthermore, a novel heterogeneous layer chaining dataflow is incorporated to reduce off-chip memory traffic stemming from extensive inter-frame motion and residual information. Thirdly, the overall architecture of NVCA is designed and synthesized in TSMC 28nm CMOS technology. Extensive experiments demonstrate that our design provides superior coding quality and up to 22.7× decoding speed improvements over other video compression designs. Meanwhile, our design achieves up to 2.2× improvements in energy efficiency compared to prior accelerators.

\end{abstract}

\section{Introduction}\label{sec:intro} 
\footnote{This work was supported in part by the National Key R\&D Program of China under Grant 2022YFB4400604. \emph{(Corresponding author: Wendong Mao, Zhongfeng Wang.)}}
Video data is rapidly growing due to the increasing demand for applications like virtual reality and surveillance. Compressing video is crucial to reducing transmission and storage costs under a limited bandwidth budget, and real-time video decoding in practical scenarios is highly desirable. During the past decades, several video coding standards, including H.264/AVC~\cite{H264}, H.265/HEVC~\cite{H265}, and H.266/VVC~\cite{H266}, have been established. However, conventional hand-crafted methods struggle to deal with complex textures and rapid motions and are less versatile for various scenarios and demands.

Recent years have witnessed the flourishing of brand-new NVC models~\cite{DVC, FVC, LU_ECCV20, ELFVC,DCVC,DBLP:conf/nips/MentzerTMCHLA22}, which integrate deep neural networks (DNNs) into hybrid video coding frameworks. Wherein the inter-frame compression, which employs the previous frames to predict the subsequent ones, has been the mainstream. Lu~\emph{et al.}~\cite{DVC} first employed two auto-encoder-style networks for NVC. Subsequently, numerous well-designed variants, including methods based on residual coding~\cite{FVC, LU_ECCV20, ELFVC}, conditional coding~\cite{DCVC}, and other technologies~\cite{DBLP:conf/nips/MentzerTMCHLA22}, have been proposed to further enhance coding efficiency. Compared to plain models~\cite{Fusion} oriented to classification or object detection, the NVC models pose new challenges for real-time deployment: \underline{1)} NVC models generally comprise multiple independent modules, such as motion compensation, feature extraction, and residual compression, causing irregular data dependencies and extensive memory access. \underline{2)} Various operations, including deconvolutions (DeConvs) and deformable convolutions (DfConvs), are incorporated into NVC models, resulting in serious \textbf{\textit{memory conflicts and computational imbalances}}. \underline{3)} NVC models are sensitive to low-bit quantization and high sparsity, bringing challenges to the lightweight designs of model compression. \underline{4)} Practical applications of NVC require \textbf{\textit{real-time processing of high-definition (HD) video streams}}, causing substantial transmission burdens. All in all, the stringent requirements for memory and processing speed underscore the significance of devising lightweight model optimizations and efficient solutions for deploying NVC models.

Despite limited researches~\cite{MobileCodec, FPXNVC} achieved deployments of NVC, they did not carry out dedicated designs and optimizations for specific hardware platforms. Moreover, many researchers focused on developing specialized accelerators~\cite{FTA, DfConv, WinoNN, 9045214} for certain operations within DNN models. However, directly applying them to NVC presents serious problems: \underline{1)} Existing accelerators struggle to handle complex data dependencies and memory access within NVC models. \underline{2)} Most accelerators mainly optimize dataflow and resource configurations for partial operations, making it difficult to simultaneously support the diverse operations within NVC models. \underline{3)} Extensive off-chip interactions for intermediate features such as motion and residual information incur significant power consumption and transmission burdens, which are unbearable for conventional accelerators. \underline{4)} Although there are currently accelerators individually supporting pruned fast convolutions (Convs)~\cite{WinoNN} or DeConvs~\cite{9045214}, the differences in the patch size and computational patterns between these two operations pose challenges in achieving simultaneous support for both. Therefore, these problems necessitate in-depth research to enhance the performance and deployability of NVC on resource-limited devices.

HD video data in RGB or YUV format is typically stored on cloud servers as encoded bitstreams. These bitstreams are then distributed to users via the Internet. To recover visualized video streams, \textit{users perform \textbf{real-time video decoding} on their mobile devices repeatedly}~\cite{ELFVC, MobileCodec}, using previously decoded frames and recently received bitstreams. In light of this, we propose an efficient accelerator for NVC, named NVCA. The main contributions of this paper are listed as follows:
\begin{itemize}
	\item At the algorithmic level, we present the CTVC-Net to improve compression quality by capturing multi-scale correlations. Additionally, we propose a fast algorithm-based sparse strategy by combining the pruning and fast algorithms in a dual approach, reducing computational complexity while maintaining compression efficiency.
	\item At the hardware level, we develop the reconfigurable sparse computing core to support sparse Convs and DeConvs based on the fast algorithm-based sparse strategy. We also incorporate a novel heterogeneous layer chaining dataflow to reduce the off-chip interactions caused by extensive inter-frame motion and residual features.
	\item The overall architecture of NVCA is meticulously developed and synthesized utilizing TSMC 28nm CMOS technology. Experimental results show that our design not only delivers superior compression performance and decoding speed over other designs but also outperforms other hardware designs in terms of energy efficiency.
\end{itemize}

\section{Background}\label{sec:bg}

\begin{figure}[hbt]
	\vspace{-0.7em}
	\centering
	\includegraphics[width=\linewidth]{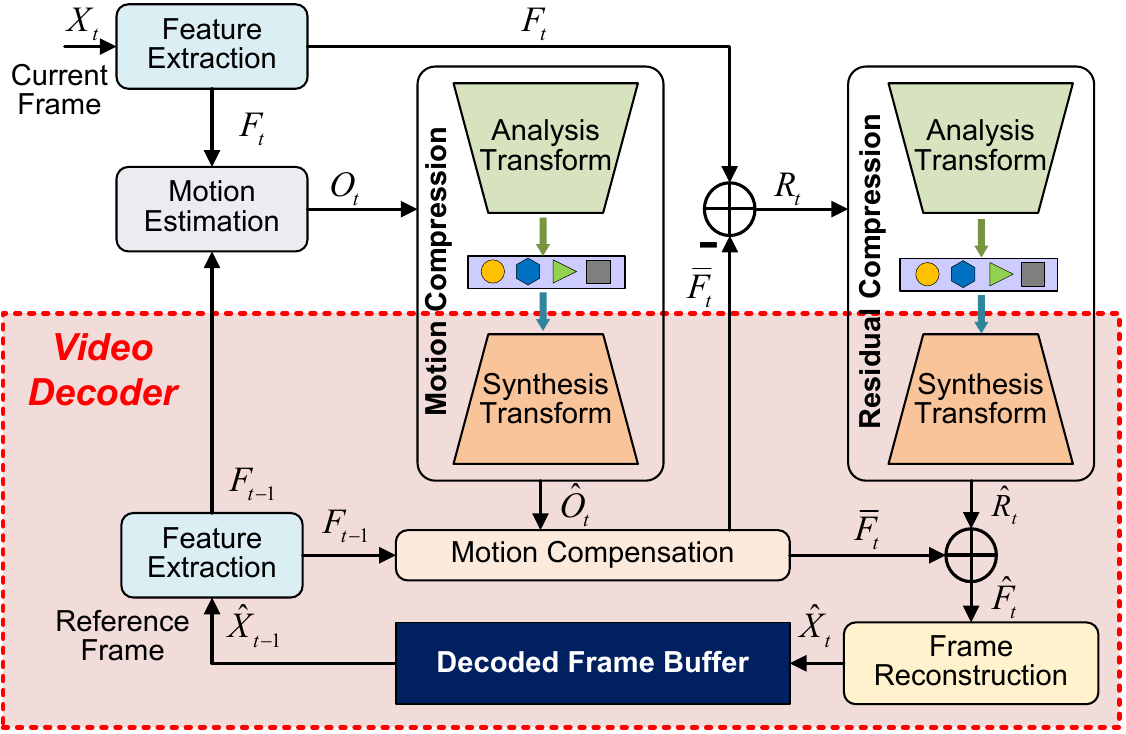}
	\vspace{-1.5em}
	\caption{Visualization of the entire process for the end-to-end NVC models. The decoding part of the NVC models is denoted by a red dashed box.}
	\label{fig:2}
	\vspace{-0.7em}
\end{figure}

NVC models follow the traditional hybrid coding frameworks. Motivated by~\cite{FVC}, all components operate within the \textit{\textbf{feature space}}, effectively reducing spatial-temporal redundancy and facilitating accurate motion estimation and compensation. As depicted in Fig.~\ref{fig:2}, the process begins with the transformation of the current frame $X_t$ and the reference frame $\hat{X} _{t-1}$ from the pixel domain to the feature domain via a \textbf{feature extraction} module. These transformed features are denoted as $F_t$ and $F_{t-1}$. To analyze the inter-frame motion relationships, a \textbf{motion estimation} module learns motion vectors $O_t$ to compress redundant temporal information between $F_t$ and $F_{t-1}$. Subsequently, $O_t$ undergoes lossy compression through an auto-encoder-style network in a \textbf{motion compression} module, resulting in the reconstructed motion $\hat{O} _{t}$. Then $\hat{O} _{t}$ is used to compensate for $F_{t-1}$ and generate the predicted feature $\bar{F}_{t}$ by a \textbf{deformable compensation} module. Following this, the spatial prediction error $R_t$ between $\bar{F}_{t}$ and $F_t$ is compressed via a \textbf{residual compression} module. The encoded motion and residual information is quantized and formed into bitstreams for transmission. Finally, by combining the reconstructed residual feature $\hat{R}_t$ with $\bar{F}_{t}$, a \textbf{feature reconstruction} module transforms the results $\hat{F}_{t}$ from the feature domain back to the pixel domain. This process results in $\hat{X}_t$, which is then stored within the decoded frame buffer. It is evident that multiple independent modules introduce complex data dependencies and extensive memory access, thereby posing significant challenges for the deployment of NVC models.

\section{Proposed Novel NVC Model}\label{sec:EDA}
\subsection{CNN-Transformer Hybrid Video Compression Network}\label{ssec:TDS}

\subsubsection{Network Structure of CTVC-Net}
\begin{figure}[hbt]
		\vspace{-1em}
	\centering
	\includegraphics[width=\linewidth]{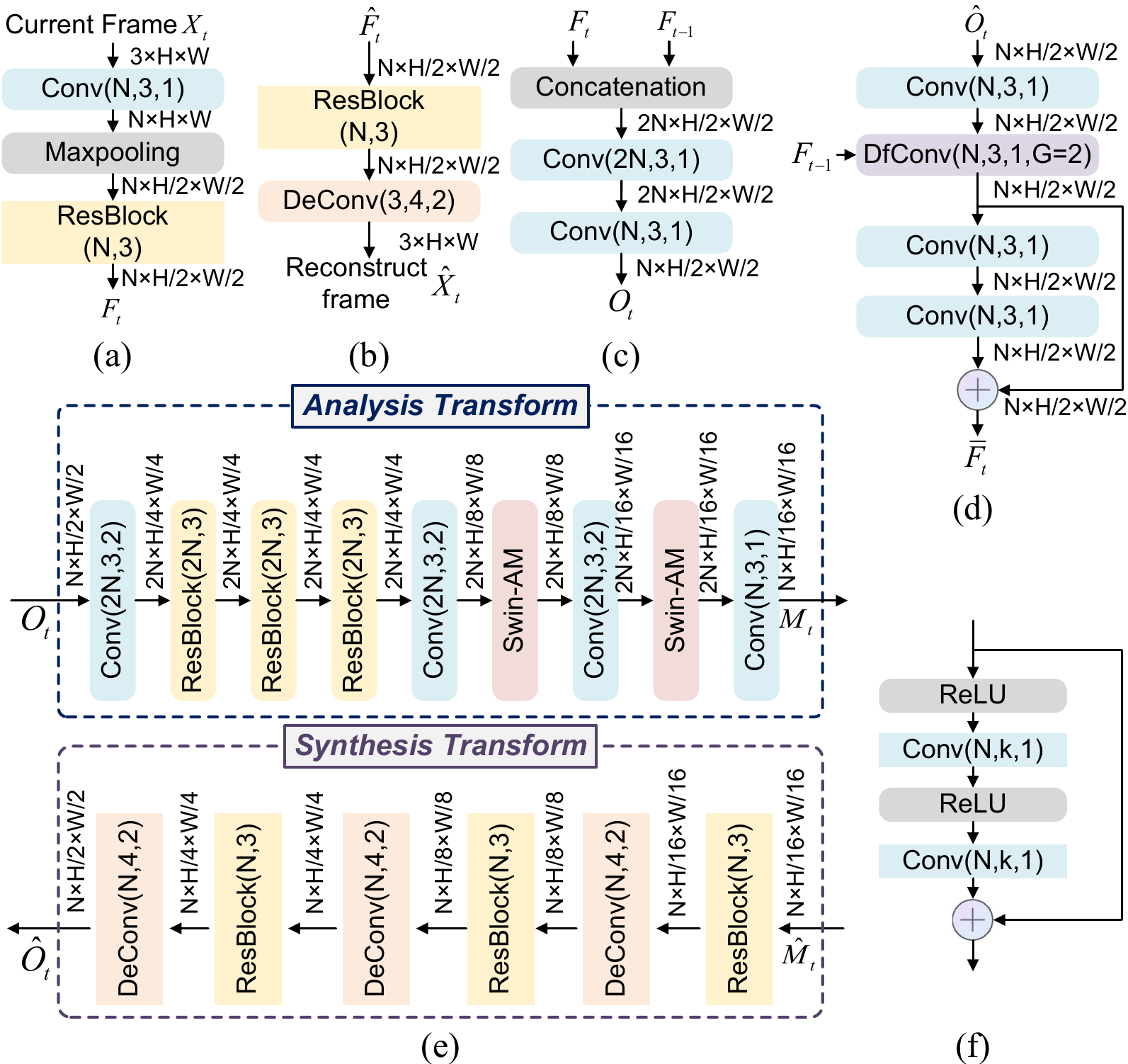}
		\vspace{-2em}
	\caption{Network structure of CTVC-Net. (a) Feature extraction. (b) Feature reconstruction. (c) Motion estimation. (d) Deformable motion compensation. (e) Motion (residual) compression. (f) Residual block (ResBlock). “Conv($N,k,s$)” and “DeConv($N,k,s$)” present the output channel number, kernel size, and stride of the Conv and DeConv, respectively. $G$ of “DfConv($N,k,s,G$)” means the group number for the DfConv.}
	\label{fig:4}
		\vspace{-1.0em}
\end{figure}

Our CTVC-Net follows the end-to-end NVC models in Fig.~\ref{fig:2}. Wherein the corresponding topologies and associated hyper-parameter settings for each module are provided in Fig.~\ref{fig:4} (a)-(e).

\subsubsection{\underline{Swin}-Transformer-based \underline{A}ttention \underline{M}odule (Swin-AM)}
\begin{figure}[hbt]
	\vspace{-1em}
	\centering
	\includegraphics[width=\linewidth]{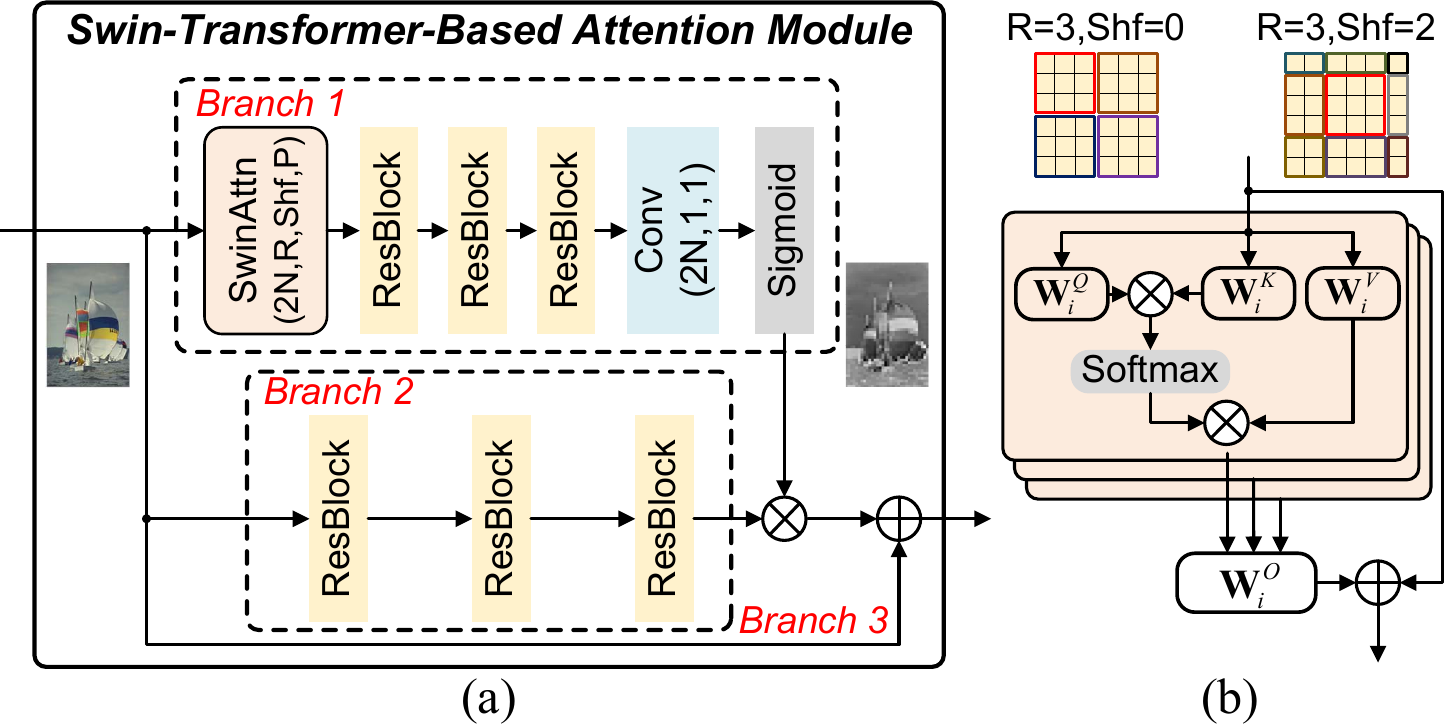}
	\vspace{-2em}
	\caption{Description of Swin-AM. (a) Network structure. ($2N,R,Shf,P$) expresses the output channel number, window size, shift value, and head number. (b) Details of SwinAtten.}
	\label{fig:5}
	\vspace{- 1.5em}
\end{figure}

Considering the strong correlations between spatially neighboring pixels, global semantic information is unsuitable for video compression~\cite{WAM}. Thus, we propose Swin-AM to enhance the compression efficiency of CTVC-Net. With the \underline{S}hift \underline{win}dow-based self-\underline{Atten}tion (SwinAtten)~\cite{SwinTrans}, Swin-AM confines the global receptive field to local non-overlapped windows while integrating the \textit{multi-scale structure information}, as depicted in Fig.~\ref{fig:5}. Specifically, Swin-AM comprises three branches. \textbf{Branch 3} manages the residual connection. \textbf{Branch 2} utilizes stacked ResBlocks to generate intermediate features. \textbf{Branch 1} employs SwinAtten and stacked Convs to mine multi-scale spatial correlations within local windows, producing window-based spatial-channel attention masks. In addition, to address the absence of feature connections across different windows, we design consecutive Swin-AMs with varying shift values $Shf$ in CTVC-Net to bridge \textit{cross-window connections}. In this way, Swin-AM effectively guides adaptive bit allocations and enhances compression efficiency by capturing local and global correlations. 

\subsection{Fast Conv and DeConv-Based Sparse Strategy}\label{ssec:DCM}
Considering NVC models' sensitivity to quantization and pruning, we seek to reduce the computational complexity of CTVC-Net by \textit{a dual approach} that combines pruning and fast algorithms. In the decoder of CTVC-Net, Convs and DeConvs play crucial roles in its high complexity. Although there are dedicated accelerators individually support pruned fast Convs or DeConvs, the intrinsic distinction between them hinders concurrent support for both. Thus, we present a fast algorithm-based united sparse strategy to reduce algorithm complexity.

\subsubsection{Fast Conv and DeConv}
Inspired by~\cite{Winograd} and~\cite{FTA}, we can express the fast Conv and DeConv by \textit{a common formula}: 
\begin{align}
	\mathbf{V}=\mathbf{A}^{\mathbf{T}}\left[\left(\mathbf{G} \mathbf{W} \mathbf{G}^{\mathbf{T}}\right) \odot\left(\mathbf{B}^{\mathbf{T}} \mathbf{X} \mathbf{B}\right)\right] \mathbf{A},
	\label{eq:1}
\end{align}
where $\mathbf{X}$, $\mathbf{W}$, and $\mathbf{V}$ denote the input patch sized at $p\times p$,  the weight sized at $k\times k$ and the output patch sized at $m\times m$, respectively. $\odot$ signifies the Hadamard product, and $\mathbf{A}$, $\mathbf{B}$ and $\mathbf{G}$ are transform matrices. Compared to individually computing each element within the output feature map, fast algorithms leverage structural similarities to generate output patches. This approach significantly reduces the computational complexity associated with both Convs and DeConvs.

To express the fast Conv and DeConv, $\mathbf{A}$, $\mathbf{B}$ and $\mathbf{G}$ can be defined by different hyper-parameters. Specifically, for the \textbf{\textit{fast Conv}} (Winograd algorithm)~\cite{Winograd} $F(m\times m, k\times k)$, the dimensions of $\mathbf{X}$ are $p\times p=(m+k-1)\times (m+k-1)$, and multiplication number is $\mu \times \mu =(m+k-1)\times (m+k-1)$. For example, the transform matrices of $F(2\times 2, 3\times 3)$:
\begin{equation}
	\resizebox{0.75\hsize}{!}{$
		\begin{split}
			\mathbf{B}_{Conv}^{\mathbf{T}} =\left[\begin{array}{cccc}
				1 & 0 & -1 & 0 \\
				0 & 1 & 1 &0 \\
				0 & -1 & 1 & 0 \\
				0 & 1 & 0 & -1
			\end{array}\right], \mathbf{G}_{Conv}=\left[\begin{array}{ccc}
				1 & 0 & 0 \\
				\frac{1}{2} & \frac{1}{2} & \frac{1}{2} \\
				\frac{1}{2} & -\frac{1}{2} & \frac{1}{2} \\
				0 & 0 & 1
			\end{array}\right],
			\label{eq:2}
		\end{split}$}
\end{equation}
\vspace{-0.4em}
\begin{equation}
	\resizebox{.45\hsize}{!}{$
	\begin{split}
		\mathbf{A}^{\mathbf{T}}_{Conv} & = \left[\begin{array}{cccc}
			1 & 1 & 1 & 0 \\
			0 & 1 & -1 & -1
		\end{array}\right].
		\label{eq:3}
	\end{split}$}
\end{equation}
It means that given a $4 \times 4$ input patch, a $3\times 3$ Conv producing a $2\times 2$ output patch requires $16$ multiplications, whereas a standard Conv needs $36$ multiplications. Regarding the \textbf{\textit{fast DeConv}} (FTA)~\cite{FTA} $T_{r}(m\times m, k\times k)$ (\textit{please \textbf{note} that it is the DeConv-tailored method and not based on Winograd algorithm}), the dimensions of $\mathbf{X}$ are $p\times p=\lceil(k+r \times s-1) / s\rceil\times \lceil(k+r \times s-1) / s\rceil$, where $m=r\times s$, $r$ represents the order. In this case, $\mu \times \mu =(k+(r-1) \times s)\times (k+(r-1) \times s)$. For example, for $T_{3}(6\times 6, 4\times 4)$ with a stride of $s=2$, the transform matrices are as follows:
\begin{equation}
	\resizebox{.894\hsize}{!}{$
		\begin{split}
			\mathbf{B}^{\mathbf{T}}_{DeConv} &\! =\!\left[\begin{array}{ccccc}
				1 & 0 & -1 & 0 & 0 \\
				0 & 1 & 1 & 0 & 0 \\
				0 & -1 & 1 & 0 & 0 \\
				0 & -1 & 0 & 1 & 0 \\
				0 & 1 & 0 & -1 & 0 \\
				0 & 0 & 1 & 1 & 0 \\
				0 & 0 & -1 & 1 & 0 \\
				0 & 0 & -1 & 0 & 1
			\end{array}\right],
			\mathbf{G}_{DeConv} & \!=\!\left[\begin{array}{cccc}
				0 & 0 & 0 & 1 \\
				0 & \frac{1}{2} & 0 & \frac{1}{2} \\
				0 & -\frac{1}{2} & 0 & \frac{1}{2} \\
				0 & 1 & 0 & 0 \\
				0 & 0 & 1 & 0 \\
				\frac{1}{2} & 0 & \frac{1}{2} & 0 \\
				-\frac{1}{2} & 0 & \frac{1}{2} & 0 \\
				1 & 0 & 0 & 0
			\end{array}\right],
			\label{eq:4}
		\end{split}$}
\end{equation}
\vspace{-1.0em}
\begin{align}
	\resizebox{.6\hsize}{!}{$
	\mathbf{A}^{\mathbf{T}}_{DeConv}=\left[\begin{array}{cccccccc}
		1 & 1 & 1 & 0 & 0 & 0 & 0 & 0 \\
		0 & 0 & 0 & 0 & 1 & 1 & 1 & 0 \\
		0 & 1 & -1 & 0 & 0 & 0 & 0 & 0 \\
		0 & 0 & 0 & 0 & 0 & 1 & -1 & 0 \\
		0 & 1 & 1 & 1 & 0 & 0 & 0 & 0 \\
		0 & 0 & 0 & 0 & 0 & 1 & 1 & 1
	\end{array}\right].
	\label{eq:5}$}
\end{align}
\subsubsection{Transform-Domain Weight Pruning}
According to Eq. \eqref{eq:1}, each element of $\mathbf{E}=\mathbf{G} \mathbf{W} \mathbf{G}^{\mathbf{T}}$ is derived by a weighted sum of uncertain elements in $\mathbf{W}$. Similarly, each element of $\mathbf{V}$ can be obtained via a weighted sum of uncertain weights $\mathbf{E}$ and inputs $\mathbf{X}$. Since the contributions of each element $\mathbf{E}_{i, j}$ to $\mathbf{V}_{c, d}$ are distinct, directly pruning $\mathbf{E}_{i, j}$ based on the magnitudes ignores their actual importance. Thus, we employ an \textit{importance factor matrix} $\mathbf{Q}$ to scale the magnitude of $\mathbf{E}_{i, j}$:
\begin{align}
	\mathbf{Q}_{i,j}=\sqrt{\sum_{\substack{0 \leqslant c, d \leqslant m-1, \\  0 \leqslant q, v \leqslant p-1}}\mathbf{H}_{c, d, i, j, q, v}^{2}}\quad 0 \leqslant i, j \leqslant \mu-1,
	\label{eq:9}
\end{align}
\vspace{-1em}
\begin{align}
	\mathbf{H}_{c, d, i, j, q, v}=\mathbf{A}_{i,c}\cdot \mathbf{A}_{j,d}\cdot \mathbf{B}_{q,i}\cdot \mathbf{B}_{v,j}.
	\label{eq:8}
\end{align}

Then, a mask matrix $\mathbf{M}$ is generated based on a predefined sparsity $\rho$ to indicate the redundant transform-domain weights:
\begin{align}
	\mathbf{M}_{i, j}=\left\{\begin{array}{ll}0 & \mathbf{Q}_{i, j}^{2} \cdot \mathbf{E}_{i, j}^{2}<\zeta\\ 1 & \mathbf{Q}_{i, j}^{2} \cdot \mathbf{E}_{i, j}^{2} \geqslant \zeta\end{array} \quad 0 \leqslant i, j \leqslant \mu-1,\right.
	\label{eq:10}
\end{align}
where $\zeta$ denotes the pruning threshold. Subsequently, the sparse Conv and DeConv can be both formulated as follows:
\begin{align}
	\mathbf{V}=\mathbf{A}^{\mathbf{T}}\left[\left(\mathbf{M}\odot \mathbf{G} \mathbf{W}\mathbf{G}^{\mathbf{T}}\right) \odot\left(\mathbf{B}^{\mathbf{T}} \mathbf{X} \mathbf{B}\right)\right] \mathbf{A}.
	\label{eq:11}
\end{align}
\vspace{-1.5em}
\section{Neural Video Compression Accelerator}
 \subsection{Overall Hardware Architecture}
\begin{figure}[hbt]
	\vspace{-0.6em}
	\centering
	\includegraphics[width=0.9\linewidth]{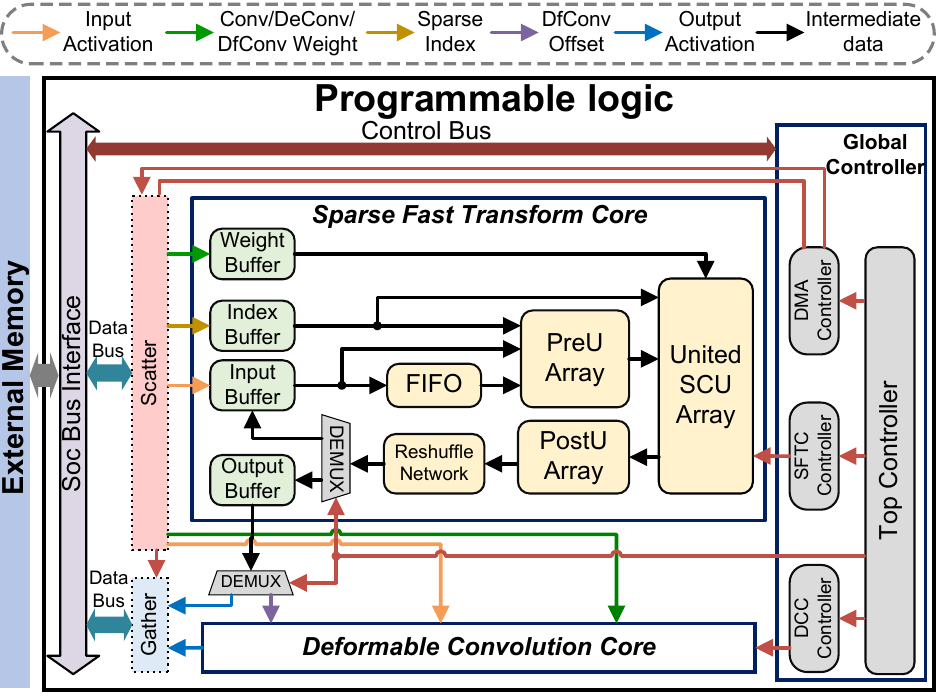}
	\vspace{-1.0em}
	\caption{Top-level block diagram of the overall architecture.}
	\label{fig:7}
	\vspace{-1.5em}
\end{figure}

In Fig.~\ref{fig:7}, the NVCA involves a Sparse Fast Transform Core (SFTC), a Deformable Convolution Core (DCC) and a global controller. The DCC, designed to support DfConvs like~\cite{DfConv}, will be briefly discussed here. The SFTC flexibly executes fast Convs or fast DeConvs by a switch flag. The computational logic involves the \textbf{Pre-processing Unit array (PreU array)}, the \textbf{Post-processing Unit array (PostU array)}, and the \textbf{united Sparse Computing Unit array (SCU array)}. The former two perform the domain transform, while the SCU array handles Hadamard products of specific data accessed by indices. On-chip memory includes the \textbf{Weight Buffer}, \textbf{Index Buffer}, \textbf{Input Buffer}, and \textbf{Output Buffer}. The Weight and Index Buffer store non-zero weights and their corresponding indices. The Input Buffer stores inputs and inter-layer features based on a novel \textit{heterogeneous layer chaining dataflow}. DeConvs' outputs are written to the Output Buffer and then transferred to the \textbf{External Memory}. Besides, \textbf{FIFO} and the \textbf{Reshuffle Network} reorder temporary input or output data before sending it to the subsequent logic.

\subsection{Sparse Fast Transform Core}
Using the proposed fast algorithm-based sparse strategy to flexibly support sparse Convs and DeConvs with minimal hardware overhead, we apply $F(2\times2, 3\times 3)$ for $3\times 3$ Conv, which carry out $16$ multiplications and $T_{3}(6\times6, 4\times 4)$ for $4\times 4$ DeConv which involves $64$ multiplications.

\subsubsection{Reconfigurable Computing Units}
\begin{figure}[hbt]
	\vspace{-0.5em}
	\centering
	\includegraphics[width=0.9\linewidth]{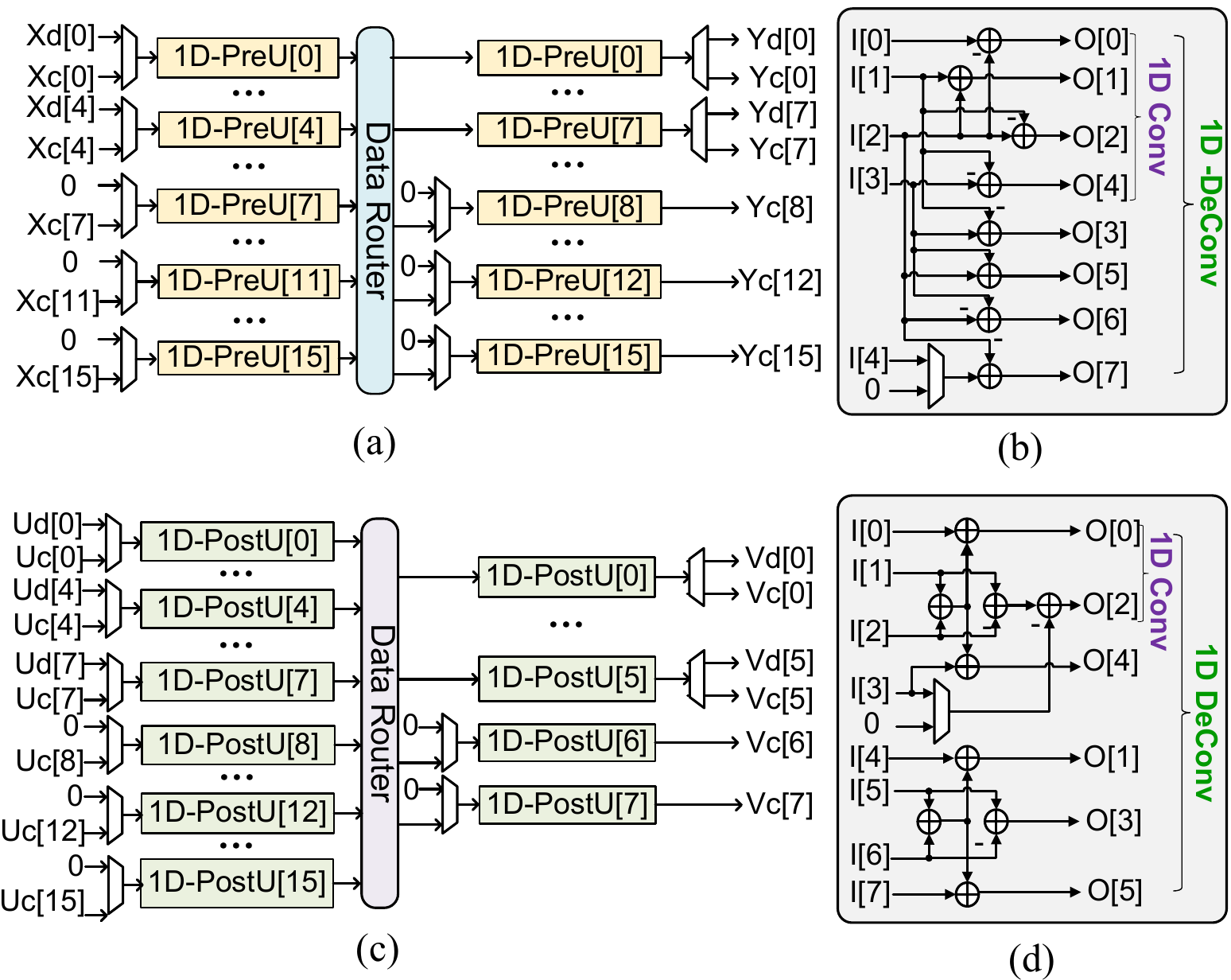}
	\vspace{-1em}
	\caption{Illustration of domain transform units. (a) The Pre-pocessing Unit (PreU). (b) 1D-PreU, which is a part of (a). (c) The Post-pocessing Unit (PostU). (d) 1D-PostU, which is a part of (c).}
	\label{fig:9}
	\vspace{-1.8em}
\end{figure}

The domain transform units of SFTC are depicted in Fig.~\ref{fig:9}. The PreU consists of $32$ 1D-PreUs, responsible for mapping input patches $\mathbf{X}$ from the spatial domain to the transform domain $\mathbf{Y} = \mathbf{B}^{\mathbf{T}} \mathbf{X} \mathbf{B}$. Similarly, the PostU, including $24$ 1D-PostUs, performs the inverse transform $\mathbf{V} = \mathbf{A}^{\mathbf{T}} \mathbf{U} \mathbf{A}$, where $\mathbf{U}$ and $\mathbf{V}$ represent the input and output patches of the PostUs. 

\begin{figure}[hbt]
	\vspace{-1em}
	\centering
	\includegraphics[width=0.97\linewidth]{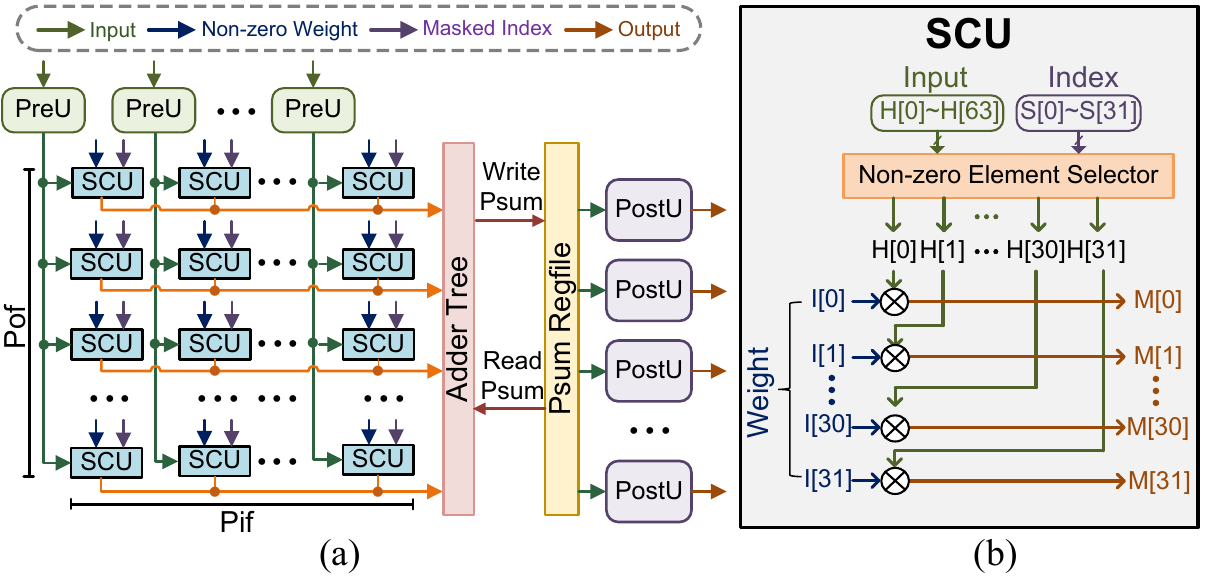}
	\vspace{-1.em}
	\caption{Computing array for sparse Convs and DeConvs. (a) The united SCU array. (b) Details of SCU, which is a part of (a).}
	\label{fig:10}
	\vspace{-0.8em}
\end{figure}

To promote data reuse in both spatial and temporal dimensions, we develop the united SCU array that flexibly supports \textit{\textbf{fine-grained structured sparsity}} while maximizing parallelism. As shown in Fig.~\ref{fig:10} (a), $Pif$ PreUs and $Pof$ PostUs are organized as a PreU array and a PostU array, respectively. The unrolling of $Pif$ input channels and $Pof$ output channels occurs along the row and column directions, forming a united SCU array of $Pif\times Pof$ SCUs, as presented in Fig.~\ref{fig:10} (b). Each SCU incorporates $64\rho$ multipliers for processing either one sparse DeConv or four sparse Convs. It selects specific inputs based on indices by a non-zero element selector, enabling element-wise multiplications with the compressed weights. After finishing the parallel Hadamard products, we reduce all input channels in the transform domain and only send the final sums to the PostU array for executing the inverse transform.

\begin{figure}[]
	\centering
	\includegraphics[width=0.9\linewidth]{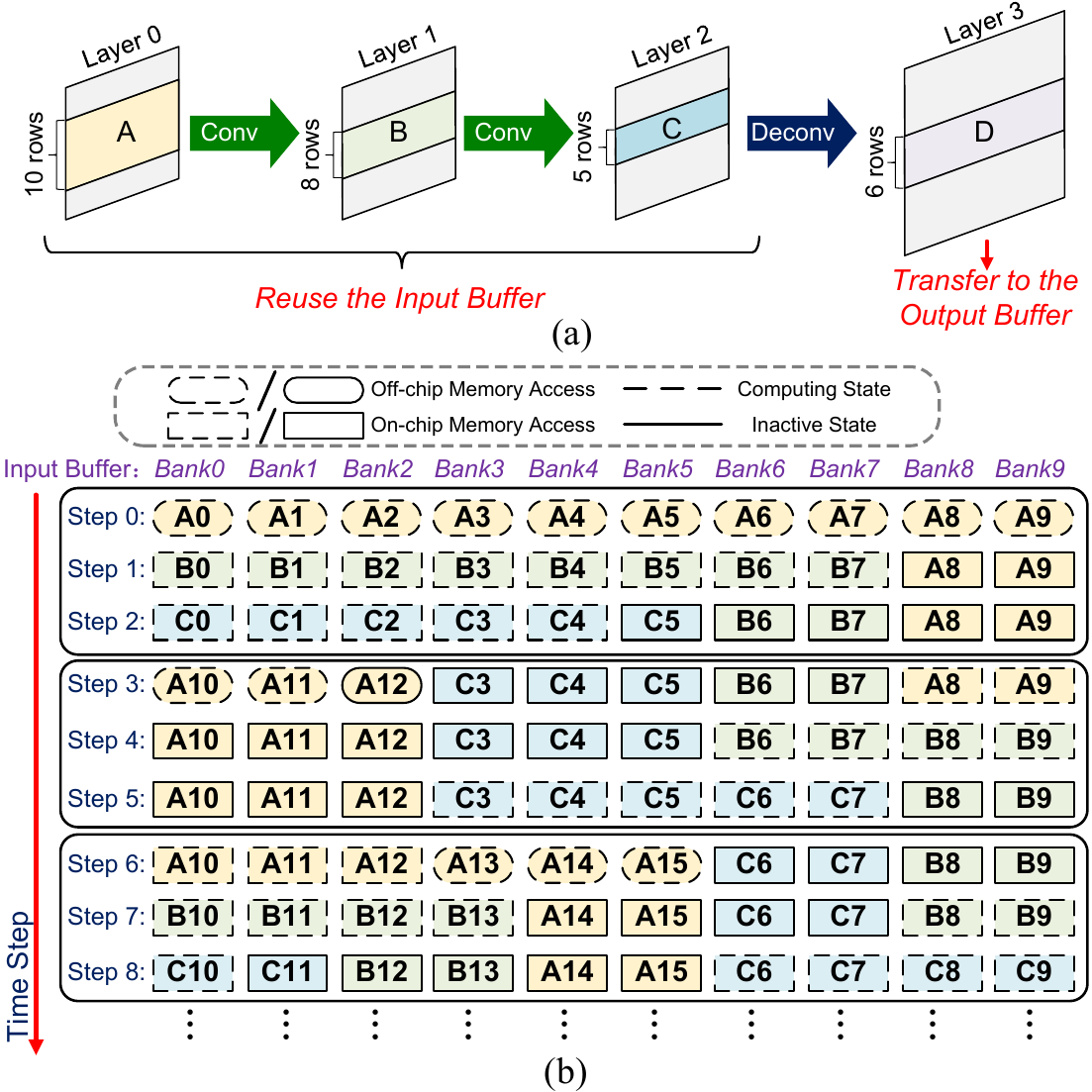}
	\vspace{-1.em}
	\caption{The heterogeneous layer chaining dataflow. Please note that the feature maps in (a) correspond directly to those in (b) in terms of color. (a) The heterogeneous layer chain is composed of Convs and DeConvs. (b) Runtime scheduling of the Input Buffer.}
	\label{fig:11}
	\vspace{-1.8em}
\end{figure}

\subsubsection{Heterogeneous Layer Chaining Dataflow}
Compressing HD videos typically involves processing and transmitting extensive motion and residual information. Additionally, multiple independent modules lead to complex dataflow and memory access patterns. All these problems incur considerable energy consumption and limit throughput due to constrained interface bandwidth. To address this issue, we propose a novel heterogeneous layer chaining dataflow to reduce massive off-chip interactions. In contrast to existing research~\cite{Fusion} of merging stacked Convs through layer fusion, the proposed method can achieve the fusion of both Convs and DeConvs while simultaneously supporting pruning and fast algorithms.

CTVC-Net decoder comprises numerous serial structures composed of two Convs followed by a DeConv, which are considered as a \textbf{chain}. In Fig.~\ref{fig:11} (a), generating the output feature map $D$ with $6$ rows requires the input feature map $A$ with $10$ rows, the intermediate feature map $B$ with $8$ rows, and the intermediate feature map $C$ with $5$ rows. To execute a \textit{\textbf{complete}} fast DeConv, the Input Buffer includes $10$ banks, denoted as Bank 0, Bank 1, and so forth. Since Conv preserves feature resolution, inter-layer features can reuse the Input Buffer through reasonable runtime scheduling. As presented in Fig.~\ref{fig:11} (b), where $Ai$ means the $i$-th row of the feature map $A$ is sent to $i\%10$-th bank. Each bank operates in either a computing or inactive state, determined by the row number required for fast Convs and DeConvs. Once some banks' data has participated in the calculations and will never be used again, these banks will be overwritten with new data from other layers. By this means, intermediate results can be stored on chip and leveraged as inputs for adjacent layers, economizing massive off-chip interactions.

\section{Evaluation} \label{sec:results}
\subsection{Experiment Setup}
We employ the Vimeo-90K dataset~\cite{Vimeo} for training, and the HEVC Class B~\cite{H265}, UVG~\cite{UVG}, and MCL-JCV~\cite{MCL-JCV} datasets for inference. Peak signal-to-noise ratio (PSNR) and multi-scale structural similarity index (MS-SSIM)~\cite{MS-SSIM} are utilized to quantitatively assess video quality, and bit per pixel (bpp) measures bit allocations for motion and residual. BDBR(\%)~\cite{FVC} indicates the average percentage of bit rate savings at the same PSNR (MS-SSIM) compared with others. In our experiments, we set hyper-parameters like $N=36$, $Pif=Pof=12$, and maintain a consistent sparsity level of $\rho=50\%$. We quantize floating-point (FP) weights and activations into fixed-point (FXP) format with 16 and 12 bits, respectively. To assess the effectiveness of hardware design, a cycle-accurate simulator~\cite{Predictor} is developed for reliable performance estimation. Additionally, we verify the simulator against RTL implementation to ensure correctness. We synthesize unit energy and area using Synopsys Design Compiler (DC) with the TSMC CMOS 28\textit{nm} HPC+ technology library.

\subsection{Algorithm Evaluation}

\begin{figure*}[]
		\vspace{-0.5 em}
	\centering
	\includegraphics[width=\linewidth]{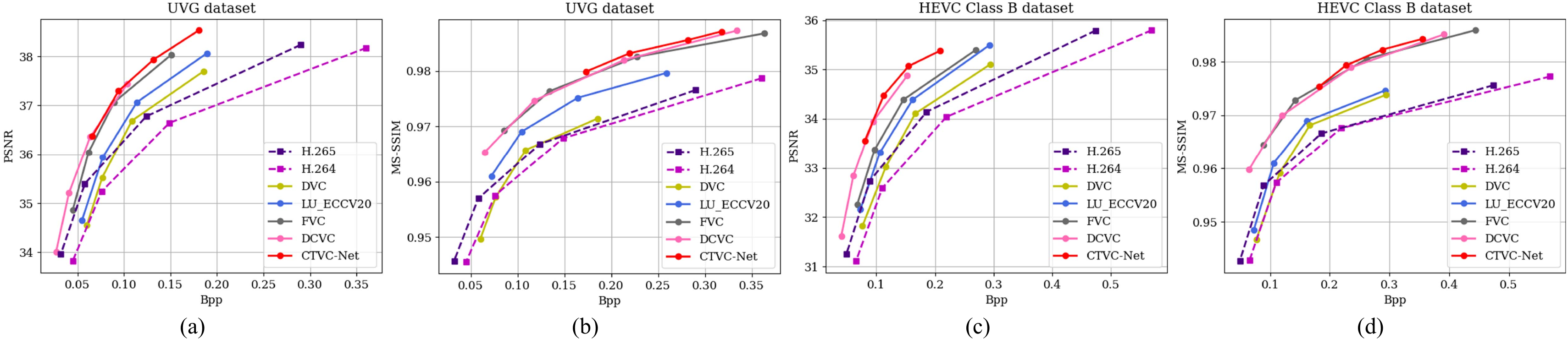}
	\vspace{-2.1 em}
	\caption{RD-curves evaluated on PSNR and MS-SSIM metrics on the UVG and HEVC Class B datasets. (a) PSNR on the UVG dataset. (b) MS-SSIM on the UVG dataset. (c) PSNR on the HEVC Class B dataset. (d) MS-SSIM on the HEVC Class B dataset.}
	\label{fig:bench_chart}
		\vspace{-2 em}
\end{figure*} 

Fig.~\ref{fig:bench_chart} illustrates the rate-distortion (RD) curves of our design on two benchmark datasets, showcasing our superior performance compared to other methods~\cite{H264,H265,DVC,FVC,LU_ECCV20,DCVC}. Our design achieves \textbf{the lowest bit consumption at the same compression quality}. Besides, in contrast to conventional hybrid video coding standards~\cite{H264,H265}, we consistently deliver the best results across all datasets. Table~\ref{tab:1} summarizes the BDBR (\%) comparisons for PSNR and MS-SSIM. We find that under 50\% sparsity, our design achieves 35.19\% and 51.30\% bit rate savings over the H.265 standard in terms of the PSNR and MS-SSIM on the UVG dataset. This reflects that by combining the Transformer-based model, our CTVC-Net effectively extracts multi-scale correlations to improve compressed video quality while significantly reducing bit rates. Besides,\textbf{ the sparse CTVC-Net maintains excellent video compression efficiency} compared to the dense version.

\begin{table}[hbt]\large
	\centering
	\vspace{-1em}
	\caption{BDBR(\%) Comparisons, Where H.265 is Employed as an Anchor.}
	\vspace{-0.5em}
	\renewcommand\arraystretch{1.2}
	\label{tab:1}
	\resizebox{\columnwidth}{!}{%
		\begin{threeparttable} 
			\begin{tabular}{c||ccc||ccc}
				\Xhline{2pt}
				& \multicolumn{3}{c||}{BDBR (PSNR) $\downarrow$} & \multicolumn{3}{c}{BDBR (MS-SSIM) $\downarrow$} \\ \cline{2-7} 
				\multirow{-2}{*}{} & \multicolumn{1}{c|}{UVG} & \multicolumn{1}{c|}{HEVC B} & MCL-JCV & \multicolumn{1}{c|}{UVG} & \multicolumn{1}{c|}{HEVC B} & MCL-JCV \\ \Xhline{1.5pt}
				H.264~\cite{H264} & \multicolumn{1}{c|}{35.27} & \multicolumn{1}{c|}{28.12} & 31.35 & \multicolumn{1}{c|}{20.06} & \multicolumn{1}{c|}{16.81} & 18.99 \\ \hline
				DVC~\cite{DVC} & \multicolumn{1}{c|}{8.45} & \multicolumn{1}{c|}{4.85} & 13.94 & \multicolumn{1}{c|}{17.29} & \multicolumn{1}{c|}{5.35} & 22.70 \\ \hline
				H.265~\cite{H265} & \multicolumn{1}{c|}{0} & \multicolumn{1}{c|}{0} & 0 & \multicolumn{1}{c|}{0} & \multicolumn{1}{c|}{0} & 0 \\ \hline
				LU\_ECCV20~\cite{LU_ECCV20} & \multicolumn{1}{c|}{-7.34} & \multicolumn{1}{c|}{-15.92} & 4.75 & \multicolumn{1}{c|}{-27.57} & \multicolumn{1}{c|}{-10.58} & 5.02 \\ \hline
				FVC~\cite{FVC} & \multicolumn{1}{c|}{-28.71} & \multicolumn{1}{c|}{-23.75} & -21.08 & \multicolumn{1}{c|}{-49.14} & \multicolumn{1}{c|}{-53.97} & {-52.45} \\ \hline
				DCVC~\cite{DCVC} & \multicolumn{1}{c|}{-35.00} & \multicolumn{1}{c|}{-37.96} & {-23.08} & \multicolumn{1}{c|}{-48.31} & \multicolumn{1}{c|}{-50.72} & -49.36 \\ \Xhline{2pt}
				\textbf{CTVC-Net(FP)} & \multicolumn{1}{c|}{-36.62} & \multicolumn{1}{c|}{-41.05} & -25.11 & \multicolumn{1}{c|}{-53.07} & \multicolumn{1}{c|}{-58.05} & -56.75 \\  \hline
				\textbf{CTVC-Net(FXP)} & \multicolumn{1}{c|}{-35.91} & \multicolumn{1}{c|}{-40.32} & -24.15 & \multicolumn{1}{c|}{-52.13} & \multicolumn{1}{c|}{-57.79} & -55.96 \\  \hline
				\textbf{CTVC-Net(Sparse)} & \multicolumn{1}{c|}{\textbf{-35.19}} & \multicolumn{1}{c|}{\textbf{-39.85}} & \textbf{-23.44} & \multicolumn{1}{c|}{\textbf{-51.30}} & \multicolumn{1}{c|}{\textbf{-57.11}} & \textbf{-55.09} \\ \Xhline{2pt}
			\end{tabular}%
			\begin{tablenotes}
				\item \textit{Negative values of BDBR mean bit rate savings, while positive values mean more bit rate consumption.}
			\end{tablenotes}
		\end{threeparttable}
	}
	\vspace{-1.2em}
\end{table}

\subsection{Hardware Evaluation}
\begin{figure}[hbt]
	\centering
	\includegraphics[width=\linewidth]{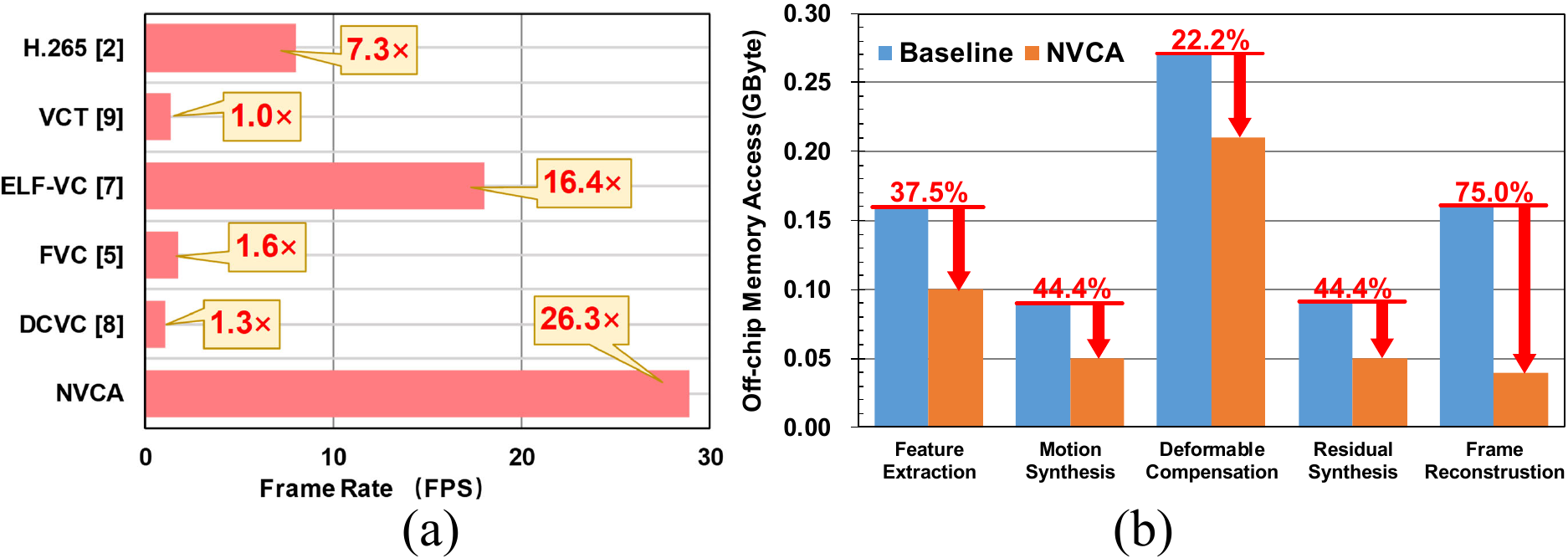}
		\vspace{-2em}
	\caption{(a) Comparison of the decoding speed. (b) Comparison of the off-chip memory access.}
	\label{fig:12}
		\vspace{-2em}
\end{figure}

To the best of our knowledge, there are \textbf{few dedicated ASIC-based accelerators for NVC}. Although some existing research effort to deploy inter-frame NVC on Snapdragon 8 chips~\cite{MobileCodec} or FPGAs~\cite{FPXNVC}, direct comparisons are difficult due to platform variations and the absence of hardware performance metrics. Thus, to verify the hardware efficiency, we compare the NVCA with a widely-used commercial GPU product (NVIDIA RTX 3090), CPU product (Intel i9-9900X), and several ASIC-based pixel processing accelerators~\cite{9585319,9570802}. In Table~\ref{tab:2}, our design achieves 2.4× higher throughput and 799.7× better energy efficiency than the GPU, and 11.1× higher throughput and 1783.9× better energy efficiency than the CPU. Moreover, we surpass~\cite{9585319,9570802} with up to\textbf{ 8.7× higher throughput and 2.2× better energy efficiency improvement}. The comparisons of average decoding time for 1080p videos between ours and other solutions~\cite{H265,DBLP:conf/nips/MentzerTMCHLA22, ELFVC, FVC, DCVC} are listed in Fig.~\ref{fig:12} (a). Notably, NVCA achieves a frame rate of 25 \textit{FPS}, outperforming DCVC~\cite{DCVC} by up to 22.7× in decoding speed. Besides, to validate the effectiveness of the heterogeneous layer chaining dataflow, we present a comparison of off-chip memory access (\textit{GByte}) for different modules in Fig.~\ref{fig:12} (b). We employ the layer-by-layer processing strategy as our baseline and achieve an overall \textbf{40.7\% reduction} in off-chip interaction compared to the baseline.

\begin{table}[hbt]\huge
	\vspace{-0.5em}
	\centering
	\caption{Comparison with Other Pixel Processing Accelerators.}
	\vspace{-0.3em}
	\renewcommand\arraystretch{1.2}
	\label{tab:2}
	\resizebox{\columnwidth}{!}{%
		\begin{threeparttable} 
			\begin{tabular}{c||c|c|c|c|c}
				\Xhline{1.5pt}
				& \multicolumn{1}{c|}{\textbf{CPU}} & \textbf{GPU} & \textbf{~\cite{9585319}} & \textbf{Alchemist~\cite{9570802}} & \textbf{NVCA} \\ \Xhline{1pt}
				\textbf{Year} & \multicolumn{1}{c|}{-} & - & 2022 & 2022 & 2023 \\ \hline
				\textbf{Task} & \multicolumn{2}{c|}{\begin{tabular}[c]{@{}c@{}}Video\\ Compression\end{tabular}} & \begin{tabular}[c]{@{}c@{}}Feature Map\\ Compression\end{tabular} & \begin{tabular}[c]{@{}c@{}}Video\\ Analysis\end{tabular} & \begin{tabular}[c]{@{}c@{}}Video\\ Compression\end{tabular} \\ \hline
				\textbf{Benchmark} & \multicolumn{1}{c|}{CTVC-Net} & CTVC-Net  & VGG16 & VGG16 & CTVC-Net \\ \hline
				\textbf{Technology (nm)} & \multicolumn{1}{c|}{14} & 8 & 28 & 65 & 28 \\ \hline
				\textbf{Frequency (MHz)} & \multicolumn{1}{c|}{3500} & 1700 & 700 & 800 & 400 \\ \hline
				\textbf{Precision (A-W)} & FP 32-32 & FP 32-32 & FXP 16-16 & FXP 16-16 & FXP 12-16 \\ \hline
				\textbf{Gate Count (M)} & \multicolumn{1}{c|}{-} & - & 1.12 & 3.03\tnote{\dag} & 5.01 \\ \hline
				\textbf{\begin{tabular}[c]{@{}c@{}}On-Chip \\ Memory (KB)\end{tabular}} & \multicolumn{1}{c|}{-} & - & 480 & 512 & 373 \\ \hline
				\textbf{Power (W)} & \multicolumn{1}{c|}{121.21} & 257.12 & 0.19 & 0.33\tnote{\dag} & 0.76 \\ \hline
				\textbf{\begin{tabular}[c]{@{}c@{}}Throughput \\ (GOPS)\end{tabular}} & \multicolumn{1}{c|}{317} & 1493 & 403 & 833 & \textbf{3525} \\ \hline
				\textbf{\begin{tabular}[c]{@{}c@{}}Energy Efficiency\\ (GOPS/W)\end{tabular}} & \multicolumn{1}{c|}{2.6} & 5.8 & 2121.1 & 2524.2 & \textbf{4638.2} \\ \Xhline{1.5pt}
			\end{tabular}
			\begin{tablenotes}
				\item \tnote{\dag} They are the scale results from 65nm technology.
			\end{tablenotes}
		\end{threeparttable}
	}
	\vspace{-0.9em}
\end{table}

\section{Conclusion} \label{sec: conclusion}
In this work, we have developed and validated an accelerator for NVC. At the algorithmic level, we propose a novel CTVC-Net and a fast algorithm-based sparse strategy to improve compression efficiency and reduce computational complexity. At the hardware level, we develop a sparse computing core and a heterogeneous layer chaining dataflow to flexibly support sparse Convs and DeConvs, significantly reducing off-chip interactions. Finally, the overall hardware architecture of NVCA is synthesized under the TSMC 28nm CMOS technology. Experimental results demonstrate that our design underscores superior improvements in video coding quality, decoding speed, and energy efficiency over prior arts.

\bibliographystyle{IEEEtran}
\bibliography{bstctl,syzhang}
\end{document}